# Direct visualization of local interaction forces in $Bi_2Sr_2CaCu_2O_{8+\delta}$ vortex matter

Aragón Sánchez J.[a][b], Cortés Maldonado R.[a], Dolz M. I.[c], Cejas Bolecek N. R.[a], van der Beek C. J.[d], Konczykowski M.[d], Fasano Y.[a][b],[1]

[a]*Laboratorio de Bajas Temperaturas, Centro Atómico Bariloche (CNEA), Av. Bustillo 9500, Bariloche 8400, Argentina*
[b]*Instituto Balseiro, Universidad Nacional de Cuyo, Av. Bustillo 9500, Bariloche 8400, Argentina*
[c]*Departamento de Física, Universidad Nacional de San Luis, Ejército de los Andes 950, San Luis 5700, Argentina*
[d]*Laboratoire des Solides Irradiés, École Polytechnique, CNRS, CEA, Université Paris-Saclay, 91128 Palaiseau, France*

**Abstract**

We study the local vortex-vortex interaction force $f_i$ of the structure frozen during a field-cooling process in an electron-irradiated $Bi_2Sr_2CaCu_2O_{8+\delta}$ sample. We compute this magnitude from snapshots of the vortex structure obtained via magnetic decoration experiments at various fields $H$ in the same sample. Since the observed structures correspond to the equilibrium ones frozen at $T \sim T_{irr}(H)$ [1], at this temperature the local modulus of $f_i$ roughly equals the local pinning force at the decorated surface of the sample. We estimate the most probable local pinning force from the mode value of the $f_i(r)$ distribution, $f_p^m$. We found that $f_p^m$ grows algebraically with $H$ and in electron-irradiated samples is 50-20% smaller than for samples with columnar defects.

## 1. Introduction

The structural properties of vortex matter in high-temperature superconductors result from the balance between thermal, pinning and vortex interaction energies. Vortices interact repulsively between them and attractively with the pinning centers naturally present or introduced in the samples. In the case of ideal samples, the stable structure is the Abrikosov perfectly hexagonal lattice. In real samples, the interaction of vortices with pinning centers introduces positional and orientational disorder that can be of short or long range depending on the nature of the pinning potential. [1] In the London limit, the vortex-vortex interaction force per unit length for a given vortex $i$ has a value

$$\boldsymbol{f}_i = \sum_j \frac{2\varepsilon_0}{\lambda} \frac{\boldsymbol{r}_{ij}}{r_{ij}} K_1\left(\frac{r_{ij}}{\lambda}\right), \quad (1)$$

Where $r_{ij}$ is the separation between the vortex $I$ and its neighbors $j$, $\varepsilon_0 = (\Phi_0/4\pi\lambda)^2$ the vortex line tension, $K_1$ the first-order modified Bessel function, and $\lambda$ the in-plane penetration depth [2]. If the vortex structure is at equilibrium, the vortex-vortex interaction force $f_i$ is compensated by the vortex-pinning interaction. Therefore, mapping the spatial distribution of $f_i(r)$ allows to estimate local pinning forces per unit length at the sample surface.[3]

Magnetic decoration is a suitable technique to perform this kind of studies since in this way the location of individual vortices at the sample surface can be imaged in large fields-of-view (thousands of vortices).[1] In a typical experiment vortices are decorated at low temperatures after field-cooling from the normal state with a slow temperature sweep-rate (tens of minutes). The decorated structure is, at lattice spacing length scales, the equilibrium one for the temperature at which the vortex positions get frozen during the field-cooling. This freezing temperature is accomplished when bulk pinning dominates and is then associated to the onset of irreversible magnetic response, namely $T_{irr}(H)$. Since this temperature strongly depends on the superconducting material and the nature of the pinning centers, a quantitative analysis requires measuring the irreversibility line for every particular sample.

1 Corresponding author. Tel.: +54-294-444-5171; fax: +54-294-444-5299.
E-mail address: yanina.fasano@cab.cnea.gov.ar

Previous works on imaging the local pinning forces via magnetic decoration have studied pnictide [4,5] and high-temperature [6] superconductors presenting rather different types of disorder (point-like versus columnar defects). Later works applied magnetic force microscopy to study the same magnitude in pnictides. [7] For the $Bi_2Sr_2CaCu_2O_{8+\delta}$ compound with disorder generated by a very dense distribution of columnar defects, the local pinning force distributions are non symmetric. [6] In these studies the disorder is correlated and strong since the irradiation with heavy ions produces amorphous columnar tracks going through the whole sample thickness. [8] In this work we study this issue in $Bi_2Sr_2CaCu_2O_{8+\delta}$ samples with point-like pinning centers resulting from the irradiation with electrons at low temperatures. [9] Electron irradiation not only modifies the pinning landscape, but also decreases the critical temperature and increases the penetration length $\lambda(0)$. [9] In order to perform a quantitative study of the local interaction forces we first measure the irreversibility line to estimate $\lambda(T_{irr}(H))$. [9]

**Experimental**

We studied a $Bi_2Sr_2CaCu_2O_{8+\delta}$ sample irradiated at 20 K with 2.3 MeV electrons with a dose of $1.7 \times 10^{19}$ el/cm$^2$ in the VINKAC facility at the École Polytechnique, France. [9] At this energy the penetration range of the electrons exceeds the sample thickness of 20 μm.[9] As a consequence of electron irradiation, the critical temperature of the sample is lowered to 66K (before irradiation Tc was of 87 K). The same sample was subsequently cleaved in order to perform field-cooling magnetic decoration experiments at various fields in fresh and flat surfaces. Magnetic decorations are performed by evaporating Fe nanoparticles that due to magnetic forces are attracted towards the vortex cores at the sample surface. For the experiments presented here this evaporation is performed at 4.2 K, after field-cooling the sample with a c-axis magnetic field following the procedure of Ref. [10]. Scanning electron microscopy is applied to image, in fields-of-view of thousands of micrometers, the clusters of Fe nanoparticles that decorate individual vortex positions. Vortex locations are digitalized to obtain $r_{ij}$. In order to visualize the local vortex-vortex interaction force, a map of typically 6000 vortices was build up from several individual pictures that partially overlap.

The zero-temperature penetration length value can be estimated from the enthalpy-jump at the first-order transition line since $\Delta B_m = 5.5\Phi_0/4\pi\lambda(0)^2$. [9] We lack information on the particular $\Delta B_m$ value for the sample we study here. However, the electron irradiation dose of our sample is slightly smaller but close to that of the sample from which the data of Fig. 4 of Ref. [9] was obtained. We estimate the penetration depth value of our sample from the data in this figure and found $\lambda(0) \sim 1.3\lambda_{pris}(0) \sim 240$ nm. This value might be an overestimation since for our sample $\Delta B_m$ is expected to be larger than that reported in Fig. 4 of Ref. [9]. Note that $\Delta B_m$ decreases on increasing the dose of electron irradiation. [9]

In order to estimate $\lambda(T_{irr}(H))$ for the evaluation of $f_i$, we apply local ac Hall magnetometry and determine the irreversibility temperature from the onset of irreversible magnetic behavior. [11] In order to detect this onset we measure $|T_{h3}|$, the normalized modulus of the third harmonic response, as described in Ref. [11]. Measurements were performed using a lock-in detection technique and applying an ac ripple field of 0.3 Oe and 177 Hz.

**Results and discussion**

Figure 1 (a) shows the temperature evolution of the third-harmonic signal $|T_{h3}|$ of the studied electron-irradiated sample for various applied fields. The onset of non-linear behavior associated to an irreversible magnetic response decreases with temperature on increasing field, see the arrows. The insert shows that the irreversibility line is roughly linear with field. This line was used to calculate $f_i(r)$ considering $\lambda(T_{irr}(H)) = \lambda(0)/\sqrt{1 - (T_{irr}/T_c)^4}$.

Figure 1 (b) shows an example of a vortex structure frozen in a field-cooling process for a 16 Gauss vortex density. White dots correspond to the vortex locations decorated with the Fe clumps evaporated at 4.2 K. The vortex positions are digitalized from sets of overlapping images like the one shown in Fig. 1 (b) for field-cooling decoration experiments at various vortex densities $B$. This information plus the $\lambda(T_{irr}(B))$ data are used to calculate the local vortex-vortex interaction force by means of the expression of Eq. 1. We considered a cutoff distance of 8 lattice parameters for the sum calculation since the contribution of the higher terms is negligible. Figure 1 (c) shows a color-coded map of the modulus of the vortex-vortex interaction force, $f_i(r)$, for a vortex density of 16 Gauss. The field-of-view indicated with a frame corresponds to the picture shown in panel (b). The spatial distribution of $f_i(r)$ does not present local patches with larger or smaller values, nor significant inhomogeneities. This is better illustrated by the probability distributions of the local values of $f_i$ shown in Fig. 1 (d) for various vortex densities: they present a distribution around a single mode value $f_i^m$. For all the studied fields the distribution of $f_i$ around the mode value is non-symmetrical (note that the data are shown in a log-log scale).

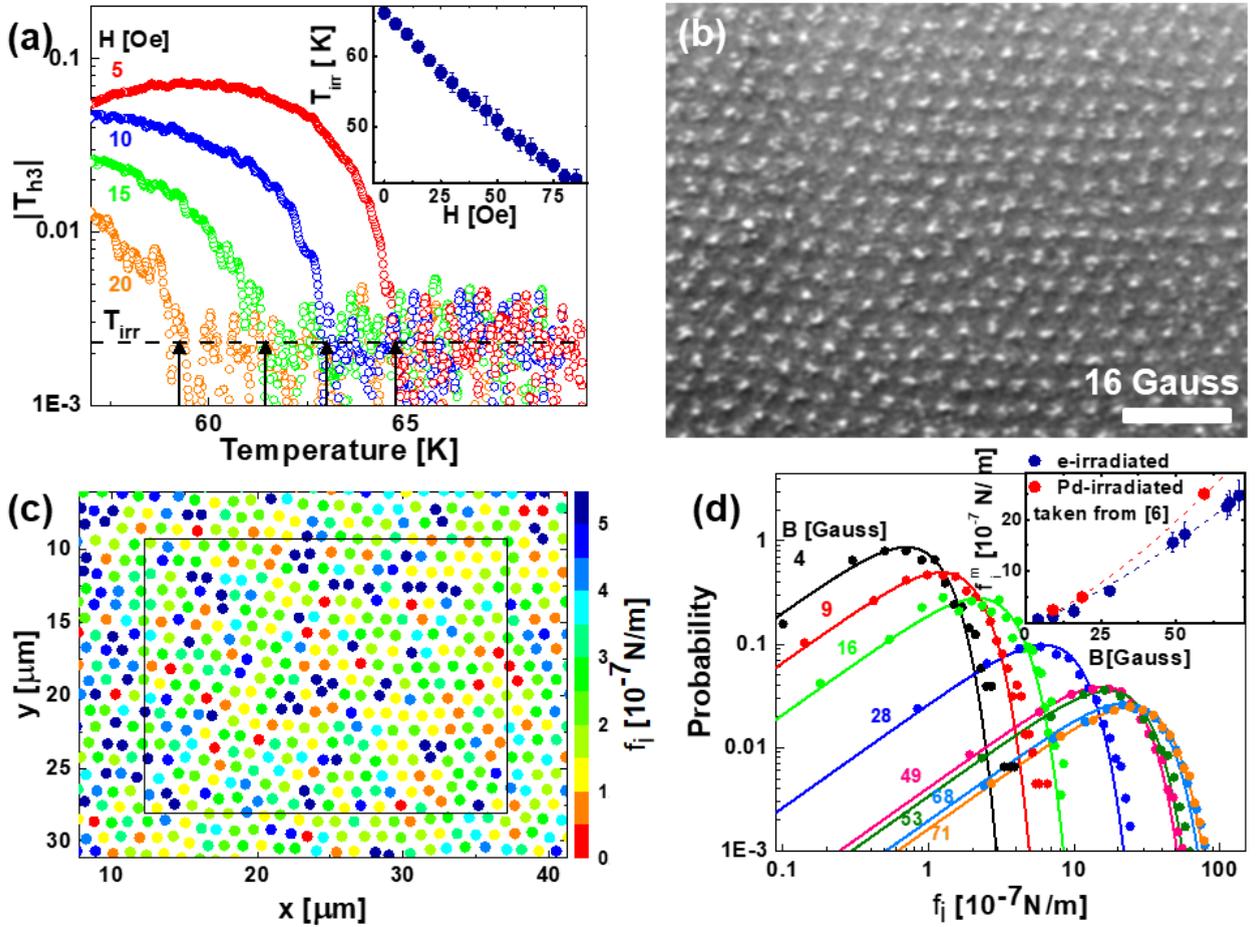

Figure 1. Local vortex-vortex interaction force and related magnitudes for vortex matter in electron-irradiated $Bi_2Sr_2CaCu_2O_{8+\delta}$ ($1.7\times10^{19}$ el/cm$^2$). (a) Modulus of the third harmonic signal of the ac magnetic response of vortex matter as a function of temperature for various applied fields. The onset of non-linear response on cooling is considered as the irreversibility temperature at which the system magnetic response becomes irreversible, $T_{irr}$. (b) Magnetic decoration image of vortex matter nucleated in a field-cooling experiment for 16 Gauss density. Vortex positions are observed as white spots; the white bar corresponds to 5 $\mu$m. (c) Color-coded map for the modulus of the vortex-vortex interaction force at 16 Gauss. The black rectangle corresponds to the image shown in (b). (d) Probability distributions of the modulus of the vortex-vortex interaction force at various fields (full points) and fits with Rayleigh functions (full lines). Insert: Field-evolution of the mode value of the interaction force distributions for the electron-irradiated sample studied here (blue points) and for the sample with columnar defects studied in Ref. [6] (red points). Dotted lines correspond to power-law fittings of the data.

However, the distribution of the $x$ and $y$ components of $\boldsymbol{f}_i$ do show a symmetric Gaussian distribution around a single mode value with a standard deviation that increases with field. This implies that the local components of the force vary at random. The modulus of a vector whose components vary following a Gaussian distribution should follow a non-symmetric, Rayleigh distribution function $(x/\sigma^2)exp(-x^2/2\sigma^2)$. The full lines of Fig. 1 (d) correspond to fits of the data with this expression. The good agreement between the fits and the data indicates that the distribution of $f_i(r)$ is quite close to a Rayleigh type and therefore the local variations of the vortex-vortex interaction force (as a vector) are at random. The mode value, the width, and the asymmetry of the $f_i$ distribution increase on enhancing vortex density. Note that for the Rayleigh distribution the standard deviation $\sigma$ is equal to the mode value $f_i^m$.

The increase of $f_i^m$ with field shown in the insert to Fig. 1 (d) is intimately linked to the increase of the standard deviation of the distribution of the force components. Moreover, $f_i^m$ growths algebraically with vortex density (see blue dotted lines). Since the vortex structure imaged by magnetic decoration corresponds to the equilibrium one frozen during the field-cooling process at $\sim T_{irr}$, the mode value of the vortex-vortex interaction force distribution can be associated to the most probable local pinning force. Then, the latter enhances algebraically with field in electron-irradiated samples, as previously observed in heavy-ion irradiated samples (see red points and lines in the insert). [6] However, the most probable pinning force for samples with a dense distribution of columnar defects is 50-20% larger than in the case of the electron-irradiated sample, the difference being 50% (20%) for low (high) fields.

**Conclusions**

In summary, the in-plane components of the vortex-vortex interaction force in $Bi_2Sr_2CaCu_2O_{8+\delta}$ samples with a high density of point defects generated by electron-irradiation vary at random for low densities of vortices with 4<$B$<71 Gauss. This phenomena was also observed for samples with a dense distribution of columnar defects generated by heavy-ion irradiation. [6] The non-zero mode values of the $f_i$ distributions come from the vortex-pinning force balancing the vortex-vortex interaction at the freezing temperature. The mode value $f_i^m$, that we interpret as the most probable pinning force, increases with field as expected. For dense distributions of pinning centers in comparison to vortex densities, this magnitude is 50-20 % larger in the case of strong columnar defect correlated-disorder than in that of weak point-disorder generated by electron irradiation.